\documentclass[a4paper]{article}
\usepackage{fullpage,cite}
\usepackage{amsmath,amsthm,amsfonts,amssymb}
\renewcommand{\tilde}{\widetilde}

\theoremstyle{remark}

\newtheorem*{rem}{Remark}
\begin{document}

\title{Quasideterminant solutions of a non-Abelian\\ Hirota-Miwa equation}
\author{C. R. Gilson, J. J. C. Nimmo\\
Department of Mathematics \\
University of Glasgow, UK\and
Y.Ohta\\
Department of Mathematics\\
Kobe University, Japan}
\date{}
\maketitle

\begin{abstract}
A non-Abelian version of the Hirota-Miwa equation is considered. In
an earlier paper [Nimmo (2006) J. Phys. A: Math. Gen. \textbf{39},
5053-5065] it was shown how solutions expressed as quasideterminants
could be constructed for this system by means of Darboux
transformations. In this paper we discuss these solutions from a
different perspective and show that the solutions are
quasi-Pl\"{u}cker coordinates and that the non-Abelian Hirota-Miwa
equation may be written as a quasi-Pl\"{u}cker relation. The special
case of the matrix Hirota-Miwa equation is also considered using a
more traditional, bilinear approach and the techniques are compared.
\end{abstract}

\section{Introduction}\label{Introduction}
There has been interest recently in noncommutative version of some
of the well-known soliton equations including the KP and KdV
equations \cite{Kupershmidt2000,Paniak2001,Sakakibara2004,
Wang2003,Wang2003a,Wang2003b,Hamanaka2003b,Hamanaka2003,Dimakis2005,Gilson2007}.
In many cases, the noncommutative version of the equation is
obtained by considering the compatibility of the same Lax pair that
is used in the commutative case. The only difference between the
commutative and noncommutative cases being that the assumption that
coefficients in the Lax pair commute is relaxed.

The Hirota-Miwa equation, or discrete KP equation,
\cite{Hirota1981,Miwa1982} is the fully discrete three dimensional
integrable system
\begin{equation}\label{hm}
        (a_2-a_3)\tau_{,1}\tau_{,23}+(a_3-a_1)\tau_{,2}\tau_{,31}
        +(a_1-a_2)\tau_{,3}\tau_{,12}=0,
\end{equation}
where $a_k$ are constants, $\tau=\tau(n_1,n_2,n_3)$ and subscripts
$_{,j}$ denote a forward shift in the corresponding discrete
variable $n_1$, $n_2$ or $n_3$. This is the compatibility conditions
for the linear system
\begin{equation}\label{eq: hm lin}
\phi_{,i}-\phi_{,j}=U_{ij}\phi,
\end{equation}
for $i<j=1,2,3$ where
\begin{equation}\label{U param}
    U_{ij}=(a_i-a_j)\frac{\tau_{,ij}\tau}{\tau_{,i}\tau_{,j}}.
\end{equation}
If one considers the linear system \eqref{eq: hm lin} in a
noncommutative setting, the parameterisation \eqref{U param} is not
appropriate but nonetheless we obtain a nonlinear system for the
$U_{ij}$ which we call the non-Abelian Hirota-Miwa equation. This
system was derived in \cite{Nimmo2006} and solutions in terms of
quasideterminants \cite{Gelfand1991} were derived by means of
Darboux transformations.

It has long been known \cite{Sato1981} that equations in the KP
hierarchy can be interpreted as Pl\"{u}cker relations on an infinite
dimensional Grassmann manifold. In this paper we establish a result
which hints at a corresponding situation in the noncommutative case
as we show that the quasideterminant solutions are, in the
terminology of Gelfand et al \cite{Gelfand1991,Gelfand2005},
quasi-Pl\"ucker coordinates and the non-Abelian Hirota-Miwa equation
is a quasi-Pl\"{u}cker relation.

This paper is arranged as follows. In Section~2, the properties of
quasideterminants that are used in the rest of the paper are
described. In Section~3, the main results from \cite{Nimmo2006} are
summarised and the quasideterminant solutions found there are
presented. The next two sections give two contrasting approaches to
direct verification of these solutions. In the first, applicable in
the general non-Abelian case, the solutions are recognised to be
quasi-Pl\"ucker coordinates and the direct verification is effected
using quasi-Pl\"ucker relations and other properties. Finally, we
consider an alternative approach which works only for the matrix
Hirota-Miwa equation. In this it is shown how standard Pl\"ucker
relations are used to verify the solutions.

\section{Preliminaries}\label{sec:prelim}

In this section we will state the properties of quasideterminants
that are used in this article. The reader is referred to the
original papers \cite{Gelfand1991,Gelfand2005} for a more detailed
description and proofs of the results described here.

\subsection{Quasideterminants}
For an $n\times n$ matrix $M=(a_{i,j})$ over a ring $\mathcal R$
(noncommutative, in general) there are $n^2$
\emph{quasideterminants} written as $|M|_{i,j}$ for $i,j=1,\dots, n$
which are also elements of $\mathcal R$. They are defined
recursively by
\begin{align}\label{quasidet}
    |M|_{i,j}&=a_{i,j}-r_i^j(M^{i,j})^{-1}c_j^i,\quad M^{-1}=(|M|_{j,i}^{-1})_{i,j=1,\dots,n}.
\end{align}
In the above $r_i^j$ represents the $i$th row of $M$ with the $j$th
element removed, $c_j^i$ the $j$th column with the $i$th element
removed and $M^{i,j}$ the submatrix obtained by removing the $i$th
row and the $j$th column from $A$. Quasideterminants can also
denoted as shown below, by boxing the entry about which the
expansion is made
\[
|M|_{i,j}=\begin{vmatrix}
    M^{i,j}&c_j^i\\
    r_i^j&\fbox{$a_{i,j}$}
    \end{vmatrix}.
\]

In the case $n=1$, let $M=(a)$,  say,  and then there is just one
quasideterminant $|M|_{1,1}=|\fbox{$a$}\;|=a$. For $n=2$, if
$M=\begin{pmatrix}a&b\\c&d\end{pmatrix}$, then there are four
quasideterminants
\begin{align*}
|M|_{1,1}=\begin{vmatrix}
    \fbox{$a$}&b\\
    c&d
    \end{vmatrix}=a-bd^{-1}c,\quad
|M|_{1,2}=\begin{vmatrix}
    a&\fbox{$b$}\\
    c&d
    \end{vmatrix}=b-ac^{-1}d,\\
|M|_{2,1}=\begin{vmatrix}
    a&b\\
    \fbox{$c$}&d
    \end{vmatrix}=c-db^{-1}a,\quad
|M|_{2,2}=\begin{vmatrix}
    a&b\\
    c&\fbox{$d$}
    \end{vmatrix}=d-ca^{-1}b.
\end{align*}
From this we can obtain the matrix inverse,
\[
    M^{-1}=\begin{pmatrix}
      (a-bd^{-1}c)^{-1}&(c-db^{-1}a)^{-1}\\
      (b-ac^{-1}d)^{-1}&(d-ca^{-1}b)^{-1}
    \end{pmatrix},
\]
which is then used in the definition of the 9 quasideterminants of a
$3\times 3$ matrix.

To explain results in the simplest way, we will sometimes write $M$
as a block matrix
\begin{equation*}\label{block}
 M=\begin{pmatrix}
      A&B\\
      C&d
    \end{pmatrix}
\end{equation*}
where $d\in\mathcal R$, $A$ is a square matrix over $\mathcal R$ of
arbitrary size and $B$, $C$ are column and row vectors over
$\mathcal R$ of compatible lengths. For this matrix,
\begin{equation}
 \begin{vmatrix}
      A&B\\
      C&\fbox{$d$}
    \end{vmatrix}
    =d-CA^{-1} B.
\end{equation}

\subsection{Commutative cases}
If the entries in $M$ commute then the above becomes the familiar
formula for the inverse of a $2\times2$ matrix with entries
expressed as ratios of determinants. Indeed this is true in general;
if the entries in $M$ commute then
\begin{equation}\label{commute}
|M|_{i,j}=\frac{\begin{vmatrix}
    M^{i,j}&c_j^i\\
    r_i^j&a_{i,j}
    \end{vmatrix}}{|M^{i,j}|}=(-1)^{i+j}\frac{|M|}{|M^{i,j}|},
\end{equation}
the alternating sign arising from the reordering of rows and columns
in the numerator.

In the special case that the ring $\mathcal R$ is the
(noncommutative) ring of $N\times N$ matrices over another
commutative ring, the $(i,j)$-quasideterminant of $M$, $|M|_{i,j}$
is itself an $N\times N$ matrix with $(k,l)$th entry
\begin{equation}\label{commute n}
(|M|_{i,j})_{k,l}=\begin{vmatrix}
    M^{i,j}&(c_j^i)_l^k\\
    (r_i^j)_k^l&\fbox{$(a_{i,j})_{k,l}$}
    \end{vmatrix}.
\end{equation}

\subsection{Invariance under row and column operations}
The quasideterminants of a matrix have invariance properties similar
to those of determinants under elementary row and column operations
applied to the matrix. Consider the following quasideterminant of an
$n\times n $ matrix;
\begin{equation}\label{invariance}
    \begin{vmatrix}
    \begin{pmatrix}
      E&0\\
      F&g
    \end{pmatrix}
    \begin{pmatrix}
      A&B\\
      C&d
    \end{pmatrix}
    \end{vmatrix}_{n,n}=
    \begin{vmatrix}
      EA&EB\\
      FA+gC&FB+gd
    \end{vmatrix}_{n,n}=g(d-CA^{-1}B)=g
    \begin{vmatrix}
      A&B\\
      C&d
    \end{vmatrix}_{n,n}.
\end{equation}
This formula describes the effect on the quasideterminants of a
matrix of performing elementary row operations, involving left
multiplication, on the matrix. The zero block in the first matrix
means that operations which add left-multiples of the row containing
the expansion point to other rows are excluded from consideration,
but all other elementary row operations are included. We see that
left-multiplying the row containing the expansion point by $g$ has
the effect of left-multiplying the quasideterminant by $g$ and that
all other operations leave the quasideterminant unchanged. There is
analogous invariance under column operations involving
multiplication on the right.

\subsection{Noncommutative Jacobi Identity}
There is a quasideterminant version of Jacobi's identity for
determinants, called the noncommutative Sylvester's Theorem  by
Gelfand and Retakh \cite{Gelfand2005}. The simplest version of this
identity is given by
\begin{equation}\label{nc syl}
    \begin{vmatrix}
      A&B&C\\
      D&f&g\\
      E&h&\fbox{$i$}
    \end{vmatrix}=
    \begin{vmatrix}
      A&C\\
      E&\fbox{$i$}
    \end{vmatrix}-
    \begin{vmatrix}
      A&B\\
      E&\fbox{$h$}
    \end{vmatrix}
    \begin{vmatrix}
      A&B\\
      D&\fbox{$f$}
    \end{vmatrix}^{-1}
    \begin{vmatrix}
      A&C\\
      D&\fbox{$g$}
    \end{vmatrix}.
\end{equation}

\subsection{Quasi-Pl\"{u}cker coordinates}
Given an $(n+k)\times n$ matrix $A$, denote the $i$th row of $A$ by
$A_i$, the submatrix of $A$ having rows with indices in a subset $I$
of $\{1,2,\dots,n+k\}$ by $A_I$ and
$A_{\{1,\dots,n+k\}\backslash\{i\}}$ by $A_{\hat\imath}$. Given
$i,j\in\{1,2,\dots,n+k\}$ and $I$ such that $\#I=n-1$ and $j\notin
I$, one defines the \emph{(right) quasi-Pl\"{u}cker coordinates}
\begin{equation}\label{rplucker}
    r^I_{ij}=r^I_{ij}(A):=
    \begin{vmatrix}
    A_I\\
    A_i
    \end{vmatrix}_{ns}
    \begin{vmatrix}
    A_I\\
    A_j
    \end{vmatrix}_{ns}^{-1}=-
    \begin{vmatrix}
    A_I&0\\
    A_i&\fbox{0}\\
    A_j&1
    \end{vmatrix},
\end{equation}
for any column index $s\in\{1,\dots,n\}$. The final equality in
\eqref{rplucker} comes from an identity of the form \eqref{nc syl}
and proves that the definition is independent of the choice of $s$.

The following are easy consequence of the definition:
\begin{align}
\label{rprop 1}
  r^I_{ij}&=0\quad\text{if $i\in I$, (and is not defined if $j$ were in $I$)}\\
\label{rprop 2}
  r^I_{ii}&=1\\
\label{rprop 3}
  r^I_{ji}&=\left(r^I_{ij}\right)^{-1}\\
\label{rprop 4}
  r^I_{ij}r^I_{jk}&=r^I_{ik}.
\end{align}

\begin{rem}
A useful consequence of \eqref{rplucker} and \eqref{rprop 3} is the
identity
\begin{equation}\label{inv}
    \begin{vmatrix}
    A^I&0\\
    A^i&\fbox{0}\\
    A^j&1
    \end{vmatrix}^{-1}=
    \begin{vmatrix}
    A^I&0\\
    A^i&1\\
    A^j&\fbox{0}
    \end{vmatrix},
\end{equation}
which shows that quasideterminants of this form may be inverted very
simply.
\end{rem}

There are also the following properties which are proved in
\cite{Gelfand2005}: \emph{skew-symmetry}
\begin{equation}
\label{rprop skew}
r^{I\backslash\{i,j\}}_{ij}r^{I\backslash\{j,k\}}_{jk}=-r^{I\backslash\{i,k\}}_{ik},
\end{equation}
for any $\{i,j,k\}\in I$ and \emph{quasi-Pl\"{u}cker relations}
\begin{equation}
\label{rprop plucker} \sum_{j\in
L}r^{L\backslash\{j\}}_{ij}r^M_{ji}=1,
\end{equation}
for any $i\notin M$, where $L,M\subset\{1,\dots,n+k\}$ such that
$\#L-1=\#M=n-1$.

\section{A non-Abelian Hirota-Miwa equation}\label{nahm}

This section is mainly a review of the results found in
\cite{Nimmo2006}. When discussing lattice equations one generally
thinks of the dependent variable as a real or complex valued
function of one or more integer-valued variables. Here we wish to
abstract this notion as far as possible and think of a lattice in a
more general way which encompasses the usual types of lattice
systems. Similar ideas were also explored by Matveev in
\cite{Matveev1998}. We suppose that dependent variables take values
in a ring $\mathcal R$ and the lattice is defined by means of
mappings $s_{i}:\mathcal R\to\mathcal R$, $i\in\mathbb N$, which we
think of as moving between neighbouring sites on the (possibly
infinite dimensional) lattice. The particular examples we have in
mind for these mappings include the usual shift operators
$s_{i}(\phi(n_i))=\phi(n_i+1)$ or $q$-shift operators
$s_{i}(\phi(n_i))=\phi(q_in_i)$, but also Darboux or B\"{a}cklund
transformations $s_{i}(\phi)=\tilde\phi$ mapping between different
solutions of the lattice equations.

With these sorts of examples in mind, we require that the $s_i$ have
the following properties:
\begin{enumerate}
\item all $s_{i}$ are invertible and mutually commute, so that motion on the lattice is
reversible and path independent
\item each $s_i$ is a linear homomorphism so that the same (linear and
nonlinear) relations satisfied at one lattice site are satisfied at
all lattice sites,
\end{enumerate}
and we refer to such mappings as \emph{lattice operators}. We will
also use the shorthand notation $X_{,i}=s_i(X)$ and
$X_{,\overline\imath}=s_i^{-1}(X)$.

Now consider lattice operators $s_i$, ($i\in\mathbb{N}$), and
invertible $U_{ij}\in\mathcal R$ and the system of linear equations
for $\phi$
\begin{equation}\label{nahm lin}
    \phi_{,i}-\phi_{,j}+U_{ij}\phi:=s_i(\phi)-s_j(\phi)+U_{ij}\phi=0,\quad
    (i,j\in\mathbb{N}).
\end{equation}
This system is algebraically over-determined and so for any $i,j,k$,
\begin{equation}
\label{nahm alg}
    U_{ij}+U_{jk}+U_{ki}=0.
\end{equation}
This includes degenerate cases which arise when some of $i,j,k$
coincide, namely $U_{ii}=0$ and $U_{ij}+U_{ji}=0$.

The compatibility conditions $\phi_{,ij}=\phi_{,ji}$ give the
nonlinear conditions
\begin{equation}\label{nahm nonlin}
U_{ij,k}U_{ik}=U_{ik,j}U_{ij}
\end{equation}
for $i,j,k$ distinct (if any of $i,j,k$ coincide this is a empty
condition) together with
\begin{equation}
\label{nahm shift}
    U_{ij,k}+U_{jk,i}+U_{ki,j}=0.
\end{equation}
By introducing $V_{ij}=U_{ij,\overline\imath\,\overline\jmath}$, one
may rewrite \eqref{nahm shift} as the ``dual'' equation
\begin{equation}
\label{nahm dual}
    V_{ij}+V_{jk}+V_{ki}=0,
\end{equation}
and one may also rewrite \eqref{nahm nonlin} in terms of $V_{ij}$ as
\begin{equation}\label{nahm nonlin dual}
V_{ij}V_{ik,\bar\jmath}=V_{ik}V_{ij,\bar k}.
\end{equation}

It is straightforward to show that the $U_{ij}$ satisfy \eqref{nahm
alg} and \eqref{nahm nonlin} if and only if the $V_{ij}$ satisfy
\eqref{nahm dual} and \eqref{nahm nonlin dual} and so it suffices to
consider either the equations for $U_{ij}$
\begin{equation}\label{nahm U}
    U_{ij}+U_{jk}+U_{ki}=0,\quad U_{ij,k}U_{ik}=U_{ik,j}U_{ij},
\end{equation}
or for $V_{ij}$
\begin{equation}\label{nahm V}
    V_{ij}+V_{jk}+V_{ki}=0,\quad V_{ik}V_{ij,\bar k}=V_{ij}V_{ik,\bar\jmath}.
\end{equation}
We call either of these equivalent systems the \emph{nonabelian
Hirota-Miwa equation}.

If we fix on one particular three dimensional sublattice (defined by
$s_1,s_2,s_3$ say), the linear equations
\begin{equation}\label{nahm lin3}
    \phi_{,i}-\phi_{,j}+U_{ij}\phi=0,\quad(i,j\in\{1,2,3\})
\end{equation}
are compatible provided \eqref{nahm U} or \eqref{nahm V} are
satisfied for $\{i,j,k\}\subset\{1,2,3\}$. The remaining dimensions
in the lattice defined by $s_4,s_5,\dots=d_1,d_2,\dots$ can be
interpreted as Darboux transformations. In \cite{Nimmo2006} these
Darboux transformations were exploited in order to construct
solutions of the non-Abelian Hirota-Miwa system expressed in terms
of quasideterminants. Finally in the section, we will state these
results.

Making the ansatz
\begin{equation}\label{ansatz}
    U_{ij}=X_{i,j}a_{ij}X_i^{-1},
\end{equation}
where $X_i\in\mathcal R$ are invertible and $a_{ij}\in\mathcal R$,
the nonlinear conditions \eqref{nahm nonlin} are satisfied for any
$X_i$ provided the $a_{ij}$ themselves satisfy this condition. We
also take $a_{ii}=0$ so that $U_{ii}=0$. With these choices,
\eqref{nahm nonlin dual} are solved by writing
\begin{equation}\label{dual ansatz}
    V_{ij}=Y_{i}^{-1}a_{ij}Y_{i,\bar\jmath},
\end{equation}
where $Y_i=X_{i,\bar\imath}^{-1}$. In the commutative case, the
skew-symmetry conditions
\[
0=U_{ij}+U_{ji}=X_{i,j}a_{ij}X_i^{-1}+X_{j,i}a_{ji}X_j^{-1}=a_{ij}\frac{X_{i,j}}{X_i}+a_{ji}\frac{X_{j,i}}{X_j},
\]
can be solved identically by the additional ansatz
$X_i=\tau_{,i}/\tau$ and taking $a_{ij}+a_{ji}=0$. The remaining
equation in \eqref{nahm U} now becomes the standard Hirota-Miwa
equation \eqref{hm} and it is for this reason that we refer to
\eqref{nahm U} and \eqref{nahm V} as non-Abelian Hirota-Miwa
equations.

Now let $X_i$, $i=1,2,3$, and $a_{ij}$ $i,j\in\{1,2,3\}$ construct,
through \eqref{ansatz}, a vacuum solution $U_{ij}$ of \eqref{nahm
lin3}. For example, one could choose $X_1=X_2=X_3=1$ and
$a_{ij}=\alpha_i-\alpha_j$. Now choose $n$ vacuum eigenfunctions,
$\theta_1,\dots,\theta_n$ i.e.~solutions of \eqref{nahm lin3} with
the vacuum choice for $U_{ij}$ made above. From these one may
construct the row vector $\theta=(\theta_1,\dots,\theta_n)$
satisfying
\begin{equation}\label{nahm lin vec}
    s_i(\theta)-s_j(\theta)+U_{ij}\theta=0.
\end{equation}
Then, after $n$ applications of Darboux transformations, the
solution obtained is $\tilde U_{ij}=\tilde X_{i,j}a_{ij}\tilde
X_i^{-1}$ where \cite{Nimmo2006}
\begin{equation}\label{nahm sol}
\tilde X_i=\begin{vmatrix}
  \theta&1\\
  s_i(\theta)&0\\
  \vdots&\vdots\\
  s_i^{n-1}(\theta)&0\\
  s_i^n(\theta)&\fbox{0}
\end{vmatrix}X_i.
\end{equation}

\section{Direct verification of solutions using quasideterminants}\label{qsol}
In this section we use the properties of quasideterminants stated in
\S\ref{sec:prelim} to effect a direct verification of the solutions
described in \eqref{nahm sol}. By recognising that the
quasideterminant in this equation is a quasi-Pl\"{u}cker coordinate
\eqref{rplucker}, it may immediately be inverted using \eqref{inv}
to obtain
\begin{equation}\label{nahm sol inv}
\tilde X_i^{-1}=X_i^{-1}\begin{vmatrix}
  \theta&\fbox{0}\\
  s_i(\theta)&0\\
  \vdots&\vdots\\
  s_i^{n-1}(\theta)&0\\
  s_i^n(\theta)&1
\end{vmatrix}.
\end{equation}
Then, use the linear equation \eqref{nahm lin vec} and the
invariance properties \eqref{invariance}, we get
\begin{equation}\label{nahm sol UV}
\tilde U_{ij}=\begin{vmatrix}
  s_j(\theta)&1\\
  s_j(s_i(\theta))&0\\
  \vdots&\vdots\\
  s_j(s_i^{n-1}(\theta))&0\\
  s_j(s_i^n(\theta))&\fbox{0}
\end{vmatrix}\begin{vmatrix}
  s_j(\theta)&\fbox{0}\\
  s_i(\theta)&0\\
  \vdots&\vdots\\
  s_i^{n-1}(\theta)&0\\
  s_i^{n}(\theta)&1
\end{vmatrix},\quad
\tilde V_{ij}=\begin{vmatrix}
  s_i^{-1}(\theta)&1\\
  \theta&0\\
  \vdots&\vdots\\
  s_i^{n-2}(\theta)&0\\
  s_i^{n-1}(\theta)&\fbox{0}
\end{vmatrix}\begin{vmatrix}
  s_i^{-1}(\theta)&\fbox{0}\\
  s_j^{-1}(\theta)&0\\
  \theta&0\\
  \vdots&\vdots\\
  s_i^{n-3}(\theta)&0\\
  s_i^{n-2}(\theta)&1
\end{vmatrix}.
\end{equation}

In fact, this is the first key step in the direct verification, we
can go one step further with $\tilde V_{ij}$. By making use of
\eqref{nahm lin vec} and \eqref{invariance} we may write all but
three rows of the quasideterminants in a way which is independent of
the particular choices of $i$ and $j$. Let $X^{(m)}$ denote
$s_k^m(X)$ for any choice of $k=1,2$ or $3$. Then it is
straightforward to verify using \eqref{nahm lin vec} and
\eqref{invariance} that
\begin{equation}\label{nahm sol V final}
\tilde V_{ij}=\begin{vmatrix}
  \theta_{,\bar\imath}&1\\
  \theta&0\\
  \vdots&\vdots\\
  \theta^{(n-3)}&0\\
  \theta^{(n-2)}&0\\
  \theta^{(n-1)}&\fbox{0}
\end{vmatrix}\begin{vmatrix}
  \theta_{,\bar\imath}&\fbox{0}\\
  \theta_{,\bar\jmath}&0\\
  \theta&0\\
  \vdots&\vdots\\
  \theta^{(n-3)}&0\\
  \theta^{(n-2)}&1
\end{vmatrix}.
\end{equation}
To complete the direct verification of these solutions we must show
that they satisfy
\begin{equation}\label{final V}
V_{ii}=0,\quad V_{ij}+V_{ji}=0,\quad V_{ij}+V_{jk}+V_{ki}=0.
\end{equation}

One consequence of \eqref{invariance} is that a quasideterminant in
which the row containing the expansion point is the same as another
row is zero. The second factor in $\tilde V_{ii}$ is such a
quasideterminant and so $\tilde V_{ii}=0$. Next define the
$(n+3)\times n$ matrix
\[
    \Theta=\begin{bmatrix}
    \theta_{,\overline1}\\
    \theta_{,\overline2}\\
    \theta_{,\overline3}\\
    \theta^{(n-2)}\\
    \theta^{(n-1)}\\
    \theta\\
    \theta^{(1)}\\
    \vdots\\
    \theta^{(n-3)}
    \end{bmatrix},
\]
which will play the role of the matrix $A$ in the definition of
quasi-Pl\"{u}cker coordinates \eqref{rplucker}, and let
$I=\{6,\dots,n+3\}$ so that
\[
    \Theta_I=
    \begin{bmatrix}
    \theta\\
    \theta^{(1)}\\
    \vdots\\
    \theta^{(n-3)}
    \end{bmatrix}.
\]
Then the quasideterminants appearing in \eqref{nahm sol V final} can
be expressed as quasi-Pl\"{u}cker coordinates,
\[
    r^{I\cup\{4\}}_{5,i}(\Theta)=-\begin{vmatrix}
  \theta&0\\
  \vdots&\vdots\\
  \theta^{(n-3)}&0\\
  \theta^{(n-2)}&0\\
  \theta^{(n-1)}&\fbox{0}\\
  \theta_{\bar\imath}&1
\end{vmatrix},\quad
r^{I\cup\{j\}}_{i,4}(\Theta)=-\begin{vmatrix}
  \theta&0\\
  \vdots&\vdots\\
  \theta^{(n-3)}&0\\
  \theta_{\bar\jmath}&0\\
  \theta_{\bar\imath}&\fbox{0}\\
  \theta^{(n-2)}&1
\end{vmatrix}.
\]
Hence,
\begin{equation}\label{nahm qpl}
    \tilde V_{ij}=r^{I\cup\{4\}}_{5,i}r^{I\cup\{j\}}_{i,4},
\end{equation}
where here and from now on we omit reference to the matrix $\Theta$.
The properties \eqref{rprop 4}, \eqref{rprop skew} and \eqref{rprop
plucker} now take the particular form
\begin{align}
\label{pp1}    r_{a,b}^{I\cup \{d\}}r_{b,c}^{I\cup \{d\}}&=r_{a,c}^{I\cup \{d\}}\\
\label{pp2}    r_{a,b}^{I\cup \{c\}}r_{b,c}^{I\cup \{a\}}&=-r_{a,c}^{I\cup \{b\}}\\
\label{pp3}    r_{a,b}^{I\cup \{c\}}r_{b,a}^{I\cup
\{d\}}+r_{a,c}^{I\cup \{b\}}r_{c,a}^{I\cup \{d\}}&=1.
\end{align}

Using \eqref{pp1} and \eqref{pp2},
\begin{align*}
\tilde V_{ij}+\tilde
V_{ji}&=r^{I\cup\{4\}}_{5,i}r^{I\cup\{j\}}_{i,4}+r^{I\cup\{4\}}_{5,j}r^{I\cup\{i\}}_{j,4}\\
&=r^{I\cup\{4\}}_{5,i}(r^{I\cup\{j\}}_{i,4}+r^{I\cup\{4\}}_{i,j}r^{I\cup\{i\}}_{j,4})\\
&=r^{I\cup\{4\}}_{5,i}(r^{I\cup\{j\}}_{i,4}-r^{I\cup\{j\}}_{i,4})=0.
\end{align*}
Similarly, using the quasi-Pl\"ucker relation \eqref{pp3} as well as
\eqref{pp1} and \eqref{pp2}, we have
\begin{align*}
\tilde V_{12}+\tilde V_{23}+\tilde
V_{31}&=r^{I\cup\{4\}}_{5,1}r^{I\cup\{2\}}_{1,4}+r^{I\cup\{4\}}_{5,2}r^{I\cup\{3\}}_{2,4}
+r^{I\cup\{4\}}_{5,3}r^{I\cup\{1\}}_{3,4}\\
&=r^{I\cup\{4\}}_{5,1}(r^{I\cup\{2\}}_{1,4}+r^{I\cup\{4\}}_{1,3}r^{I\cup\{4\}}_{3,2}r^{I\cup\{3\}}_{2,4}
+r^{I\cup\{4\}}_{1,3}r^{I\cup\{1\}}_{3,4})\\
&=r^{I\cup\{4\}}_{5,1}(r^{I\cup\{2\}}_{1,4}-r^{I\cup\{4\}}_{1,3}r^{I\cup\{2\}}_{3,4}
-r^{I\cup\{3\}}_{1,4})\\
&=r^{I\cup\{4\}}_{5,1}(1-r^{I\cup\{4\}}_{1,3}r^{I\cup\{2\}}_{3,1}
-r^{I\cup\{3\}}_{1,4}r^{I\cup\{2\}}_{4,1})r^{I\cup\{2\}}_{1,4}=0,
\end{align*}
and so the verification is complete.

This verification gives a suggestion that quasi-Pl\"ucker relations
play a similarly important role in direct verification of solutions
do in the noncommutative situation as standard Pl\"ucker relations
do in the commutative case and we intend to pursue this idea more
fully in future work.

\section{An alternative method for the matrix Hirota-Miwa equation}
In the case that $\mathcal R$ is the ring of $r\times r$ complex
matrices, an alternative, more familiar direct approach is
available. This involves expressing the matrix entries as ratios of
determinants, reexpressing the system \eqref{nahm V} in bilinear
form and verifying the solutions using identities arising from
Laplace expansions. In this section, we will contrast this more
familiar, bilinear approach with the direct verification using
quasideterminants described in the last section. We will see that
the proof a similar sequence of steps but, in contrast with the
proof involving quasideterminants, has a rather \emph{ad hoc} feel
to it, in which the details of the calculation such as keeping track
of the correct power of $-1$, are rather intricate.

In the case that we seek solutions $V_{ij}$ of \eqref{nahm V} has
solutions in this matrix ring, the ingredients in the solutions
\eqref{nahm sol V final}, the elements $\theta_k$ are themselves
also (invertible) $r\times r$ matrices and the concatenation of
these, $\theta$ is a $r \times rn$ matrix. This expression for the
solutions \eqref{nahm sol V final} as quasideterminants can then be
reexpressed in terms of a product of matrices whose entries are
ratios of $n\times n$ determinants,
\begin{equation}\label{factor}
V_{ij}=\frac{G_i}{F_i}\; \frac{h_{ij}}{F_j},
\end{equation}
where for each $i,j$, $F_i$ is scalar and $G_i$ and $h_{ij}$ are $r
\times r$ matrices. For a more compact notation, allowing
determinants to be expressed in terms of columns rather than rows,
$\theta$ is replaced by its transpose so that now $\theta$ is a $rn
\times r$ matrix. The specific expressions are
\begin{equation}\label{Fi}
F_i=
\begin{vmatrix}
 s^{-1}_i (\theta) &
  \theta &
\theta^{(1)}& \cdots& \theta^{(n-2)}
\end{vmatrix},
\end{equation}
and the $(p,q)$th entries of $G_i$ and $h_{ij}$ are
\begin{equation}\label{Gi}
(G_i)_{pq}=(-1)^{rn+1+q}
\begin{vmatrix} s^{-1}_i (\theta)_{\hat q}
&
  \theta &
\theta^{(1)}& \cdots& \theta^{(n-2)}& \theta^{(n-1)}_p
\end{vmatrix},
\end{equation}
and
\begin{equation}\label{hij}
(h_{ij})_{pq}=(-1)^{r+q}
\begin{vmatrix}
s^{-1}_i (\theta)_{p} &
 s^{-1}_j(\theta) &
  \theta &
\cdots& \theta^{(n-3)}& \theta^{(n-2)}_{{\hat q}}
\end{vmatrix},
\end{equation}
respectively, where $\theta_p$ denotes the $p^{\text{th}}$ column of
$\theta$ and $\theta_{\hat q}$ means all the columns of $\theta$
except the $q^{\text{th}}$.

If $V_{ij}$ is to satisfy \eqref{final V} then we must show that
\begin{equation}\label{bilin}
    h_{ii}=0,\quad G_ih_{ij}+G_jh_{ji}=0,\quad
    G_ih_{ij}F_k+G_jh_{jk}F_i+G_kh_{ki}F_j=0.
\end{equation}
The first of these is obvious since if $j=i$ then two columns in
$h_{ii}$ are the same. We will now show how the other two conditions
may be verified.

For each $p,q$ consider the $2rn\times2rn$ determinants
\begin{equation*}
L_{ij}=\left|
\begin{array}{ccccc|ccccc|c}
\theta & \theta^{(1)} & \cdots& \theta^{(n-2)}&  \theta^{(n-1)}_p &0
&0&\cdots& 0& 0  &  s^{-1}_i (\theta)     \cr \hline 0&0&\cdots &
0&0& s^{-1}_j(\theta) &\theta & \cdots& \theta^{(n-3)} &
\theta^{(n-2)}_{{\hat q}}& s^{-1}_i(\theta)
\end{array}
\right|.
\end{equation*}
The expansion of $L_{ij}$ by $rn\times rn$ minors is, up to a sign
independent of $i$ and $j$,
\begin{align*}
\sum_{m=1}^r (G_i)_{pm} (h_{ij})_{mq}=(G_ih_{ij})_{pq}.
\end{align*}
By using elementary row and column operations we also see that
$L_{ji}=-L_{ij}$, and so we have verified the second conditions
\begin{equation}\label{id1}
    G_ih_{ij}+G_jh_{ji}=0.
\end{equation}

For the final condition, it is necessary to introduce auxiliary
matrix variables $J_{ij}$ where
\begin{equation}\label{J}
 (J_{ij})_{pq}= (-1)^q
 \begin{vmatrix}
s^{-1}_i(\theta)_{{\hat q}}&  s^{-1}_j (\theta)_p & \theta & \cdots&
\theta^{(n-2)}
\end{vmatrix}.
\end{equation}
By a similar expansion, the determinants
\begin{equation*}\label{bigd3}
M_{ij}=\left|
\begin{array}{cccc|c|cccc}
s^{-1}_i(\theta)_{{\hat q}} & \theta & \cdots& \theta^{(n-2)}&
s^{-1}_j (\theta)  &0 &\cdots& 0 &0    \cr \hline 0&0&\cdots & 0&
s^{-1}_j(\theta) &\theta & \cdots& \theta^{(n-2)} & \theta^{(n-1)}_{
p}
\end{array}
\right|=(-1)^q(G_jJ_{ij})_{pq}.
\end{equation*}
This determinant can also be rewritten as
\begin{equation*}\label{bigd4}
(-1)^{rn+1}\left|
\begin{array}{cccc|c|cccc}
s^{-1}_i(\theta)_{{\hat q}} & \theta & \cdots& \theta^{(n-2)}&
\theta^{(n-1)}_{ p} &0  &0 &\cdots& 0    \cr \hline 0&0&\cdots & 0&
\theta^{(n-1)}_{ p}& s^{-1}_j(\theta)&\theta & \cdots&
\theta^{(n-2)}
\end{array}
\right|=(-1)^q(G_iF_j)_{pq}.
\end{equation*}
So we get
\begin{equation}\label{id2}
G_j J_{ij}=G_i F_j,
\end{equation}
for any $i,j$.

Finally, consider expansion of the determinants
\begin{align*}\label{bigd5}
&\left|
\begin{array}{ccccc|ccc|cccc}
s^{-1}_i\theta & \theta & \cdots&
\theta^{(n-3)}&\theta^{(n-2)}_{\hat q} &  s^{-1}_k \theta &(s^{-1}_j
\theta)_{p} & (\theta^{(n-2)})_q   &0 &\cdots&0& 0    \cr \hline
0&0&\cdots & 0&0& s^{-1}_k \theta &(s^{-1}_j \theta)_{p} &
(\theta^{(n-2)})_q   & \theta & \cdots&
\theta^{(n-3)}&\theta^{(n-2)}_{\hat q}
\end{array}
\right|\\
&\qquad=(-J_{kj} h_{ki} +h_{ji} F_k  -F_i h_{jk})_{pq}.
\end{align*}
After elementary row and column operations, this determinant can
also be show to vanish and so we get
\begin{equation}\label{id3}
-J_{kj} h_{ki} +h_{ji} F_k  -h_{jk}F_i=0.
\end{equation}

Using \eqref{id1} and \eqref{id2} and then \eqref{id3} it follows
that
\begin{align*}
G_ih_{ij}F_k+G_jh_{jk}F_i+G_kh_{ki}F_j&=G_j(-h_{ji}F_k +h_{jk} F_i
+J_{kj}h_{ki})=0.
\end{align*}
This completes the alternative direct verification in the matrix
case.

\section{Conclusions}

In this paper we have considered two direct approaches to verifying
quasideterminant solutions of a non-Abelian Hirota-Miwa equation
found in \cite{Nimmo2006}. In the first approach, the solutions are
expressed in terms of quasi-Pl\"ucker coordinates and the
verification is achieved by using properties of these objects,
including quasi-Pl\"ucker relations. The second approach, available
only in the matrix case and not involving quasideterminants, uses an
expression for the solutions as matrix of ratios of determinants. In
the first approach, the machinery used is more sophisticated but the
verification is quite straightforward and applicable whatever the
nature of the non-Abelian equation, matrix or quaternion or
otherwise. The second verification needs less sophisticated tools
(transformation to a bilinear form and standard Pl\"ucker relations)
and is not always available since determinants are not defined in
the general non-Abelian case. A direct quasideterminant approach has
also been applied to noncommutative versions of the KP and mKP
equations \cite{Gilson2007,Gilson2007a} and it is hoped that this
can be developed into a more widely applied direct approach to
noncommutative integrable systems.

\end{document}